\documentclass[12pt,a4paper]{article}\parindent=6.32631mm
\usepackage{amsfonts,amsmath,braket,cite,float,graphicx,mathrsfs,xcolor}
\newcommand\jm{@tecnico.ulisboa.pt}\newcommand\jz{\mathbb{Z}}\newcommand\jx
{\times}\newcommand\jf[1]{\footnote{#1}}\newcommand\jzs{\jz_2\jx\jz'_2}
\newcommand\jzb{\jz_2}\newcommand\jzc{\jz_3}\newcommand\js{\vspace{5mm}}
\newcommand\jphi[2]{\big(\phi^\dagger_{#1}\phi_{#2}\big)}

\newcommand{\mM}[1]{M_{#1}}\newcommand{\mMd}{\mM{d}}\newcommand{\mMu}{\mM{u}}
\newcommand{\wM}[1]{M_{#1}^0}\newcommand{\wMd}{\wM{d}}\newcommand{\wMu}{\wM{u}}
\newcommand{\wMdag}[1]{M_{#1}^{0\dagger}}
\newcommand{\rM}[1]{\hat M_{#1}^0}\newcommand{\rMd}{\rM{d}}\newcommand{\rMu}{\rM{u}}
\newcommand{\nvec}[2]{\hat n_{#1}^{#2}}

\newcommand{\refeq}[1]{eq.\eqref{#1}}
\newcommand{\refeqs}[2]{eqs.\eqref{#1}--\eqref{#2}}

\begin{document}\begin{titlepage}\hfill\begin{minipage}[r]{0.3\textwidth}
\begin{flushright}CFTP/20-005\\IFIC/20-20\end{flushright}\end{minipage}
\begin{center}\js{\large\bf Extending Trinity to the Scalar Sector through 
Discrete Flavoured Symmetries}\\\js Jo\~{a}o M. Alves$^{a,}$\jf{\tt j.maga%
lhaes.alves\jm}, Francisco J. Botella$^{b,}$\jf{\tt francisco.j.botella@uv%
.es}, Gustavo C. Branco$^{a,}$\jf{\tt gbranco\jm} and Miguel Nebot$^{a,}$%
\jf{\tt miguel.r.nebot.gomez\jm}\end{center}\js\begin{flushleft}\emph{$^a$ 
Departamento de F\'{i}sica and Centro de F\'{i}sica Te\'{o}rica de Part\'{i
}culas (CFTP),\\\quad Instituto Superior T\'{e}cnico (IST), U. de Lisboa (%
UL),\\\quad Av. Rovisco Pais 1, P-1049-001 Lisboa, Portugal.}\\\emph{$^b$ %
Departament de F\'{i}sica Te\`{o}rica and Instituto de F\'{i}sica Corpuscu%
lar (IFIC),\\\quad Universitat de Val\`{e}ncia -- CSIC, E-46100 Valencia, %
Spain.}\end{flushleft}\js

\begin{abstract}
 We conjecture the existence of a relation between elementary scalars and fermions, making it plausible the existence of three Higgs doublets. We introduce a Trinity Principle (TP) which, given the fact that there are no massless quarks, requires the existence of a minimum of three Higgs doublets. The TP states that each line of the mass matrix of a quark of a given charge should receive the contribution from one and only one scalar doublet and furthermore a given scalar doublet should contribute to one and only one line of the mass matrix of a quark of a given charge. This principle is analogous to the Natural Flavour Conservation (NFC) of Glashow and Weinberg with the key distinction that NFC required the introduction of a flavour blind symmetry, while the TP requires a flavoured symmetry, to be implemented in a natural way. We provide two examples which satisfy the Trinity Principle based on $\mathbb{Z}_3$ and $\mathbb{Z}_2\times\mathbb{Z}_2'$ flavoured symmetries, and show that they are the minimal multi-Higgs extensions of the Standard Model where CP can be imposed as a symmetry of the full Lagrangian and broken by the vacuum, without requiring soft-breaking terms. We show that the vacuum phases are sufficient to generate a complex CKM matrix, in agreement with experiment. The above mentioned flavoured symmetries lead to a strong reduction in the number of parameters in the Yukawa interactions, enabling a control of the Scalar Flavour Changing Neutral Couplings (SFCNC). We analyse some of the other physical implications of the two models, including an estimate of the enhancement of the Baryon Asymmetry of the Universe provided by the new sources of CP violation, and a discussion of the the strength of their tree-level SFCNC.
\end{abstract}

\end{titlepage}

\section{Introduction\label{SEC:intro}}In the Standard Model (SM), all masses are generated 
by the vacuum expectation value of a single Higgs doublet. Since there is %
no fundamental reason for having the minimal scalar sector when the fermio%
n space is non-trivial, multi-Higgs models can arise in a variety of well-%
motivated scenarios (see \cite{Branco:2011iw,Ivanov:2017dad} and references therein). For instance, the first Two Higgs Doublet Model (2HDM) was introduced by Lee \cite{Lee:1973iz} to achieve spontaneous CP Violation (CPV).

In general, multi-Higgs extensions of the SM generate large Scalar Flavour 
Changing Neutral Couplings (SFCNC) described by many parameters. Glashow an%
d Weinberg \cite{Glashow:1976nt} have shown that such problems can be avoi%
ded in 2HDMs by introducing a flavour blind $\jzb$ symmetry which constrai%
ns each fermion of a given charge to couple with only one scalar doublet. %
As such, their model implements Natural Flavour Conservation (NFC) despite 
having two Higgs fields. Alternatively, Branco, Grimus and Lavoura (BGL)  %
\cite{Branco:1996bq} introduced a symmetry in 2HDMs to generate tree-level 
SFCNC completely determined by the CKM matrix $V$. In some of those models%
, there is a significant suppression of the most dangerous SFCNC. As an ex%
ample, the $K^0-\bar{K}^0$ transition becomes proportional to $(V_{td}V^*_{
ts})^2$ in the BGL models of type ``top''. BGL models were extended to the leptonic secto%
r \cite{Botella:2011ne} and their physical signals have been extensively a%
nalysed in the literature \cite{Botella:2009pq,Botella:2011ne,Botella:2014ska,Bhattacharyya:2014nja,Botella:2015hoa,Bednyakov:2018hfq}. In \cite{Alves:2017xmk,Alves:2018kjr}, they were generalised in a sc%
heme where symmetries are introduced to reduce the free parameters in a 2HDM%
. The new models are distinguished from a BGL model by the presence of simultane%
ous tree-level SFCNC in both quark sectors.

One of the fundamental questions in Particle Physics is the origin of the %
triplication of fermion families. In this paper, we conjecture that a rela%
tion between scalar doublets and fermions exists, making it plausible to c%
onsider three copies of the former. To avoid a proliferation of flavour pa%
rameters, we introduce the Trinity Principle (TP) which is analogous to th%
e NFC of Glashow and Weinberg:
\begin{quote}
``Each line of the mass matrix of a quark of a given charge should receive the contribution from one and only one scalar doublet and furthermore a given scalar doublet should contribute to one and only one line of the mass matrix of a quark of a given charge.''
\end{quote}
Since all quarks have a non-vanishing mass, it is clear that the TP demands a minimum of three Higgs doublets to be implemented. 
Notice that no Three Higgs Doublets Model (3HDM) satisfies simultaneously the TP and NFC \cite{Grossman:1994jb,Cree:2011uy}, since they require each quark
of a given charge to couple with either three or only one scalar, respectively. 
Furthermore, the TP predictions are only stable under renormalization when a symmetry of the full Lagrangian is introduced, with soft breaking terms in its scalar potential allowed nonetheless.

At this stage, it is worth recalling the various attempts at constructing %
viable models of spontaneous CPV, in the context of multi-Higgs extensions 
of the SM. Note that in a model with spontaneous CPV Yukawa couplings are %
real, so that the vacuum phases have to be able to generate a complex CKM %
matrix, since experiment has shown that the CKM is complex even if one all%
ows for the presence of physics beyond the SM \cite{Botella:2005fc,Charles:2004jd,Bona:2005vz}. I%
n the Lee model, one can have spontaneous CPV and in the presence of three 
quark families generate a complex CKM matrix from the vacuum phase. Howeve%
r in the Lee model there are SFCNC which are not under control. In order t%
o solve the problem of SFCNC Glashow and Weinberg suggested the NFC princi%
ple implemented through a flavour-blind $\jzb$ symmetry, which makes it im%
possible to have either explicit or spontaneous CPV in the scalar sector, %
unless the discrete symmetry is softly broken \cite{Branco:1985aq}. 
 Even though spontaneous CPV can be achieved with NFC \cite{Weinberg:1976hu,Branco:1980sz}, this leads to mass matrices with only a complex global phase which can be removed via a rephasing of the right-handed fermions, resulting in a real CKM matrix. 
Recently, a minimal model was proposed with a flavoured softly broken $\jzb$ symmetry, where CP is spontaneously violated and the vacuum phase is able to %
generate a complex CKM matrix \cite{Nebot:2018nqn}.

In this paper, we construct two 3HDMs with flavoured symmetries which implement the TP. Both models have the notable featu%
re of being the minimal multi-Higgs extensions of the SM invariant under an exact symmetry of the Lagrangian where spontaneous CPV can produce a complex CKM matrix.

This paper is organised as follows: in the next two sections, we review th%
e Yukawa interactions of a 3HDM before parametrizing the SFCNC of both mod%
els. In sections \ref{SEC:CPscalar} and \ref{SEC:CKM}, the CP properties of their scalar sectors are de%
rived, followed by the generation of a complex CKM matrix out of the vacuu%
m phases. Some of the physical implications of these models are discussed in section \ref{SEC:imp}, such as the strength of their tree-level SFCNC, the size of the Electric Dipole Moment (EDM) of the neutron and the enhancement of the Baryon Asymmetry of the Universe (BAU). Finally, we provide our conclusions in the last section.

\section{The 3HDM: Generalities and Notation\label{SEC:gen}}We settle the notation by rev%
iewing the quark Yukawa interactions of a general 3HDM,
\begin{equation}\label{eq:LqY}
\mathscr{L}_{qY}=-\bar{q}^0_L\big(\phi_1\Gamma_1+\phi_2\Gamma_2+\phi_3
\Gamma_3\big)d^0_R-\bar{q}^0_L\big(\tilde{\phi}_1\Delta_1+\tilde{\phi}_2
\Delta_2+\tilde{\phi}_3\Delta_3\big)u^0_R+h.c.,
\end{equation}
where $\tilde{
\phi}_a=i\sigma_2\phi^*_a$ and all omitted flavour indices are summed over.
After spontaneously breaking the electroweak symmetry, we introduce the Hi%
ggs basis with $\braket{\phi_a}\propto v_ae^{i\alpha_a}$, $v^2=v^2_1+v^2_2+
v^2_3$, $v'^2=v^2_1+v^2_2$, $v''=vv'/v_3$ and $x=-v'^2/v_3$,
\begin{equation}\begin{pmatrix}H_1\\H_2\\H_3\end{pmatrix}=\begin{pmatrix}v_
1/v&v_2/v&v_3/v\\v_2/v'&-v_1/v'&0\\v_1/v''&v_2/v''&x/v''\end{pmatrix}\begin
{pmatrix}e^{-i\alpha_1}\phi_1\\e^{-i\alpha_2}\phi_2\\e^{-i\alpha_3}\phi_3
\end{pmatrix},\end{equation}where the would-be Goldstone bosons $G^+$ and $
G^0$ are identified in\begin{equation}H_1=\begin{pmatrix}G^+\\\frac{1}{
\sqrt{2}}(v+H^0+iG^0)\end{pmatrix},\quad H_2=\begin{pmatrix}C^+\\\frac{1}{
\sqrt{2}}(R+iI)\end{pmatrix},\quad H_3=\begin{pmatrix}C'^+\\\frac{1}{\sqrt{
2}}(R'+iI')\end{pmatrix}.\end{equation}
Introducing $\Gamma'_a=e^{i\alpha_a}\Gamma_a$ and $\Delta'_a=e^{-i\alpha_a}\Delta_a$, one can read the fermion mass matrices
\begin{equation}
\wMd=\frac{1}{\sqrt{2}}\left(v_1\Gamma'_1+v_2\Gamma'_2+v_3\Gamma'_3\right),\quad 
\wMu=\frac{1}{\sqrt{2}}\left(v_1\Delta'_1+v_2\Delta'_2+v_3\Delta'_3\right),
\end{equation}
which are diagonalized by the unitary transformations\footnote{Notice that one can assume, without loss of generality, the correct ordering of the quark masses in the diagonalization: a permutation $s\in S_3$ can be represented by $P_s\in O(3)$ such that $P_s^T\mathrm{diag}\{m_1,m_2,m_3\}P_s=\mathrm{diag}\{m_{s(1)},m_{s(2)},m_{s(3)}\}$ with $m_{s(1)}<m_{s(2)}<m_{s(3)}$;  $P_s$ can then be absorbed in the diagonalizing $U_{fX}$ matrices with $U_{fX}\mapsto P_s U_{fX}$.} of the fermion fields $f^0_L=U_{fL}f_L$ and $f^0_R=U_{fR}f_R$,
\begin{equation}
\mMd=U^\dagger_{dL}\wMd U_{dR}=\mathrm{diag}\{m_d,m_s,m_b\},\quad
\mMu=U^\dagger_{uL}\wMu U_{uR}=\mathrm{diag}\{m_u,m_c,m_t\}.
\end{equation}
The CKM mixing matrix is $V\equiv U^\dagger_{uL}U_{dL}$. Similarly, the Yukawa couplings of $H_2$ and $H_3$ in the fermion mass bases are given by
\begin{equation}\label{eq:Nq:1}
\begin{aligned}
&N_d=\frac{1}{\sqrt{2}}U^\dagger_{dL}\left(v_2\Gamma'_1-v_1\Gamma'_2\right)U_{dR},\quad
&&N'_d=\frac{1}{\sqrt{2}}U^\dagger_{dL}\left(v_1\Gamma'_1+v_2\Gamma'_2+x\Gamma'_3\right)U_{dR},\\
&N_u=\frac{1}{\sqrt{2}}U^\dagger_{uL}\left(v_2\Delta'_1-v_1\Delta'_2\right)U_{uR},\quad
&&N'_u=\frac{1}{\sqrt{2}}U^\dagger_{uL}\left(v_1\Delta'_1+v_2\Delta'_2+x\Delta'_3\right)U_{uR}.
\end{aligned}
\end{equation}
(In the initial weak basis, these Yukawa couplings are $N^{(')0}_f=U_{fL}N^{(')}_fU^\dagger_{fR}$).
The couplings between neutral scalars and fermions in a general 3HDM follow from
\begin{equation}\label{eq:YukNeutral:0}
\mathscr{L}_{qY}\supset-\sum_{f=u,d} \bar f_{L}\left[H^0\mM{f}/v+(R+i\epsilon_fI)N_f/v'+(R'+i\epsilon_fI')N'_f/v''\right]f_{R}+\mathrm{h.c.}\,,
\end{equation}
where $\epsilon_d=-\epsilon_u=1$ and implicit generation indices are summed over. Notice that these are not yet the Yukawa couplings of physical neutral scalars, since $\{H^0,R,R',I,I'\}$ are not mass eigenstates.

\section{Implementations of the Trinity Principle\label{SEC:TP}}We put forward the TP to 
be imposed on multi-Higgs models analogously to the NFC of Glashow and Wei%
nberg. The two are distinguished by the number of Higgs doublets which mus%
t couple to a quark of a given charge\jf{Three in the TP versus one in NFC%
.}. Thus, the former is implemented by flavoured symmetries while the latt%
er requires flavour-blind constructions.\\ 

One can expect a limited number of possible implementations of the TP: since each doublet should only contribute to one and only one line of the mass matrix, and each line corresponds to the couplings with one left-handed quark doublet, that can be accomplished in three different manners\footnote{We recall that there is no intrinsic meaning attached to the labels $1,2,3$ of the different quark and scalar generations in \refeq{eq:LqY}.}.
\begin{enumerate}
\item Each scalar doublet $\phi_j$ couples to the same quark doublet $q_{Lj}^0$ in both quark sectors.
\item One scalar doublet, for example $\phi_1$, couples to the same quark doublet $q_{L1}^0$ in both quark sectors while the remaining $\phi_j$ couple to different $q_{Lk}^0$ in the two quark sectors, for example $\phi_2$ couples to $\bar q_{L2}d_R$ and $\bar q_{L3}u_R$, while $\phi_3$ couples to $\bar q_{L3}d_R$ and $\bar q_{L2}u_R$.
\item Each scalar doublet $\phi_j$ couples to different quark doublets $q_{Lj}^0,q_{Lk}^0$ in the two quark sectors, for example $\phi_1$ couples to $\bar q_{L1}d_R$ and $\bar q_{L2}u_R$, $\phi_2$ couples to $\bar q_{L2}d_R$ and $\bar q_{L3}u_R$, and $\phi_3$ couples to $\bar q_{L3}d_R$ and $\bar q_{L1}u_R$.
\end{enumerate}
The first and second possibilities appear to be achievable in terms of a symmetry (see appendix \ref{sAPP:implem:sym} for details), and we concentrate on them in the following.
In this section, we study the flavour sectors of these two implementations of the TP. We start with case 2 above: the model is invariant under a $\jzc$ transformation, and has some similarity with the so-called ``jBGL'' models introduced in \cite{Alves:2018kjr}, which are, however, 2HDMs rather than the 3HDMs discussed here. Then we address case 1 above: the model is invariant under a $\jzs$ symmetry and has some similarity with the so-called ``gBGL'' models introduced in \cite{Alves:2017xmk} (which are, again, 2HDMs rather than 3HDMs). It is to be noted that the latter contains, as particular cases, the extensions of BGL models to the 3HDM context discussed in \cite{Botella:2009pq}.

\subsection{$\jzc$ model -- flavour structure\label{sSEC:TP:Z3flavour}}In the $\jzc$ model, the TP %
is implemented through the following symmetry,
\begin{equation}\label{eq:Z3sym}
\begin{aligned}
& \phi_1\to \phi_1,\quad && \phi_2\to \Upsilon\,\phi_2,\quad && \phi_3\to \Upsilon^{-1}\,\phi_3,\\
& q^0_{L1}\to q^0_{L1},\quad && q^0_{L2}\to \Upsilon\,q^0_{L2},\quad && q^0_{L3}\to \Upsilon^{-1}\,q^0_{L3}, 
\end{aligned}
\end{equation}
with all other fields transforming trivially and $\Upsilon=\exp(2\pi i/3)$. 
From this expression, we obtain the Yukawa couplings
\begin{equation}\label{eq:Yuk:Z3}
\begin{split}\Gamma_1=\begin
{pmatrix}\jx&\jx&\jx\\0&0&0\\0&0&0\end{pmatrix},\quad\:\Gamma_2&=\begin
{pmatrix}0&0&0\\\jx&\jx&\jx\\0&0&0\end{pmatrix},\quad\:\Gamma_3=\begin
{pmatrix}0&0&0\\0&0&0\\\jx&\jx&\jx\end{pmatrix},\\\Delta_1=\begin{pmatrix}
\jx&\jx&\jx\\0&0&0\\0&0&0\end{pmatrix},\quad\Delta_2&=\begin{pmatrix}0&0&0
\\0&0&0\\\jx&\jx&\jx\end{pmatrix},\quad\Delta_3=\begin{pmatrix}0&0&0\\\jx&
\jx&\jx\\0&0&0\end{pmatrix},
\end{split}\end{equation}
with $\jx$ an arbitrary complex number. By using these textures on \refeq{eq:Nq:1}, we can derive
\begin{equation}\label{eq:Nq:Z3}
\begin{aligned}
&N_d=\left(v_2/v_1P^{dL}_1-v_1/v_2P^{dL}_2\right)\mMd,\quad && N'_d=\left(1-v^2/v^2_3P^{dL}_3\right)\mMd,\\
&N_u=\left(v_2/v_1P^{uL}_1-v_1/v_2P^{uL}_3\right)\mMu,\quad && N'_u=\left(1-v^2/v^2_3P^{uL}_2\right)\mMu,
\end{aligned}
\end{equation}
where we introduce the projection operators $P^X_i=U_X^\dagger P_iU_X$ with $(P_i)_{jk}=\delta_{ij}\delta_{ik}$. 
The projection operators obey completeness identities $\sum_i P^X_i=\mathbf{1}$, and, by construction, $P^{uL}_iV=VP^{dL}_i$. It follows that, in addition to vacuum expectation values, fermion masses and elements of the CKM matrix $V$ -- all of them fixed in other sectors of the model --, two projectors $P^X_i$ are sufficient to describe \refeq{eq:Nq:Z3}, and thus all the physical Yukawa couplings. Let us consider the question in detail. First, one can identify the rows of $U_X$ with complex orthonormal vectors $\nvec{i}{X}$ with components $[\nvec{i}{X}]_j\equiv [U_X]_{ij}$ \cite{Alves:2017xmk,Alves:2018kjr}, that is
\begin{equation}
U_{dL}=\begin{pmatrix}\leftarrow & \nvec{1}{d} & \rightarrow\\ \leftarrow & \nvec{2}{d} & \rightarrow\\ \leftarrow & \nvec{3}{d} & \rightarrow\end{pmatrix},\quad
U_{uL}=\begin{pmatrix}\leftarrow & \nvec{1}{u} & \rightarrow\\ \leftarrow & \nvec{2}{u} & \rightarrow\\ \leftarrow & \nvec{3}{u} & \rightarrow\end{pmatrix},
\end{equation}
Then, by construction,
\begin{equation}
(P^X_{i})_{jk}=(U_X)_{ij}^\ast(U_X)_{ik}=[\nvec{i}{X}]_j^\ast[\nvec{i}{X}]_k\,,\qquad [\nvec{i}{u}]_j\, V_{jk}=[\nvec{i}{d}]_k\,.
\end{equation}
 Choosing one projector $P^X_{i_1}$, 5 real parameters are in general required to describe $\nvec{i_1}{X}$, but a global rephasing of $\nvec{i_1}{X}$ leaves $(P^X_{i_1})_{jk}$ unchanged\footnote{Equivalently, $U_X$ and $U_{\rm Ph}U_X$ give the same projectors $P^X_j$ for $U_{\rm Ph}=\text{diag}\{e^{i\varphi_1},e^{i\varphi_2},e^{i\varphi_3}\}$.}, and thus only 4 real parameters are required to describe the appearance of $P^X_{i_1}$ in \refeq{eq:Nq:Z3}. Next, one needs a second projector $P^X_{i_2}$, $i_2\neq i_1$; unitarity of $U_X$ implies that the corresponding complex unit vector $\nvec{i_2}{X}$ is orthogonal to $\nvec{i_1}{X}$, i.e. $\sum_j [\nvec{i_2}{X}]_j^\ast[\nvec{i_1}{X}]_j=0$, and thus only 2 additional real parameters are required for $P^X_{i_2}$. Overall, all the Yukawa couplings in \refeq{eq:Nq:Z3} depend on vacuum expectation values, fermion masses, elements of the CKM matrix $V$, and, in general, the 6 new real parameters describing the projectors $P^X_{i_1}$, $P^X_{i_2}$. Notice in particular how the appearance of the fermion mass factors controls the intensity of SFCNC.\\ 

Notice that we are yet to provide any WB independent definition of these models. 
In appendix \ref{sAPP:implem:wb}, we show that the relation below can play such a role,
\begin{equation}\label{eq:WB:Z3}
\Gamma^\dagger_i\Gamma_{j\neq i}=\Gamma_1^
\dagger\Delta_{i\neq1}=\Gamma^\dagger_2\Delta_{i\neq3}=\Gamma^\dagger_3
\Delta_{i\neq2}=0,\quad\Gamma_i\neq0.
\end{equation}

\subsection{$\jzs$ model -- flavour structure\label{sSEC:TP:Z2Z2flavour}}
We construct the $\jzs$ model by requiring invariance under the transformations
\begin{equation}\label{eq:Z2Z2sym}
\jzb:~\:
\phi_1\rightarrow-\phi_1,\quad{q^0_L}_1\rightarrow-{q^0_L}_1,\quad\quad\jz'
_2:~\:\phi_2\rightarrow-\phi_2,\quad{q^0_L}_2\rightarrow-{q^0_L}_2,
\end{equation}
with all other fields transforming trivially under them. Thu%
s, its Yukawa couplings are
\begin{equation}\label{eq:Yuk:Z2Z2}
\begin{split}\Gamma_1=\begin
{pmatrix}\jx&\jx&\jx\\0&0&0\\0&0&0\end{pmatrix},\quad\:\Gamma_2&=\begin
{pmatrix}0&0&0\\\jx&\jx&\jx\\0&0&0\end{pmatrix},\quad\:\Gamma_3=\begin
{pmatrix}0&0&0\\0&0&0\\\jx&\jx&\jx\end{pmatrix},\\\Delta_1=\begin{pmatrix}
\jx&\jx&\jx\\0&0&0\\0&0&0\end{pmatrix},\quad\Delta_2&=\begin{pmatrix}0&0&0
\\\jx&\jx&\jx\\0&0&0\end{pmatrix},\quad\Delta_3=\begin{pmatrix}0&0&0\\0&0&0
\\\jx&\jx&\jx\end{pmatrix}.
\end{split}\end{equation}
It is now straightforward to evaluate the tree-level SFCNC of a $\jzs$ model as
\begin{equation}\label{eq:Nq:Z2Z2}
\begin{aligned}
&N_d=\left(v_2/v_1P^{dL}_1-v_1/v_2P^{dL}_2\right)\mMd,\quad && N'_d=\left(1-v^2/v^2_3P^{dL}_3\right)\mMd,\\
&N_u=\left(v_2/v_1P^{uL}_1-v_1/v_2P^{uL}_2\right)\mMu,\quad && N'_u=\left(1-v^2/v^2_3P^{uL}_3\right)\mMu.
\end{aligned}
\end{equation}
As before, the flavour structure of this model can be entirely described with two projectors $P^X_i$. Thus, both implementations of the TP can be described with 6 real parameters.
We finish this section with a WB independent definition for the $\jzs$ model, namely
\begin{equation}\label{eq:WB:Z2Z2}
\Gamma^\dagger_i\Gamma_j=\Gamma^\dagger_i\Delta_j=0,\quad\Gamma_i\neq0,
\quad\quad i\neq j.
\end{equation}

\section{CPV in the Scalar Sector\label{SEC:CPscalar}}
In this section we study the CP properties of the scalar sector of the two models with flavoured symmetries. It has been pointed out \cite{Branco:2015bfb} that the introduction of a symmetry $S$ in the scalar sector of a multiple Higgs doublet model, has consequences for CP violation.  In all examples of two- and three-Higgs-doublet models with symmetries, one observes the following remarkable property: if $S$ prevents explicit CP-violation (CPV), at least in the neutral Higgs sector, then it also prevents spontaneous CPV, and if $S$ allows explicit CPV, then it also allows for spontaneous CPV. In reference \cite{Branco:2015bfb} it was  conjectured that this is a general phenomenon and it was proven that the conjecture holds for any rephasing symmetry group $S$ and for any number of doublets. In our analysis we will confirm that this conjecture indeed holds for the case of the two models with flavoured symmetries.

\subsection{$\jzc$ model -- explicit and spontaneous CPV\label{sSEC:CPscalar:Z3}}The most general %
scalar potential of a 3HDM invariant under \refeq{eq:Z3sym} is given by
\begin{equation}\label{eq:potV:Z3}
\begin{split}V(\phi)&=\mu^2_{ii}\jphi{i}{i}+\lambda_i\jphi{
i}{i}^2+\lambda_{ij}\jphi{i}{i}\jphi{j}{j}+\lambda'_{ij}\jphi{i}{j}\jphi{j%
}{i}\\&+\Big[\sigma_1\jphi{1}{2}\jphi{1}{3}+\sigma_2\jphi{2}{1}\jphi{2}{3}+
\sigma_3\jphi{3}{1}\jphi{3}{2}+h.c.\Big],
\end{split}\end{equation}
where there is an implicit sum over $i<j=1,2,3$. While by hermiticity the paramete%
rs $\mu^2_{ii}$, $\lambda_i$, $\lambda_{ij}$ and $\lambda'_{ij}$ are real, 
the coefficients $\sigma_1$, $\sigma_2$ and $\sigma_3$ remain complex.

By considering a CP transformation $\phi^{CP}_a=e^{i\gamma_a}\phi^*_a$, it 
is straightforward to verify that CP is explicitly broken in this potential
unless the product $\sigma_1\sigma_2\sigma_3$ is real.

When CP is imposed as a symmetry of the Lagrangian, the scalar fields can %
always be rephased to select $\sigma_1$, $\sigma_2$ and $\sigma_3$ real.
 The terms in \refeq{eq:potV:Z3} sensitive to the vacuum phases read
\begin{equation}\label{eq:VvevZ3}
V(\braket{\phi})\supset\frac
{1}{2}v_1v_2v_3\big[\rho_1\cos\big(\alpha_++\alpha_-\big)+\rho_2\cos\alpha_
++\rho_3\cos\alpha_-\big],
\end{equation}
with $\alpha_+=\alpha_3+\alpha_1-2
\alpha_2$, $\alpha_-=\alpha_1+\alpha_2-2\alpha_3$ and $\rho_i=\sigma_iv_i$.
By minimizing this potential with respect to $\alpha_+$ and $\alpha_-$, we 
find a trivial CP-conserving solution $\sin\alpha_+=\sin\alpha_-=0$, as well as the conditions for a CP-violating vacuum
\begin{equation}\label{eq:SCPVvac}
\cos\alpha_+=\frac{\rho_1\rho_
3}{2\rho^2_2}-\frac{\rho_1}{2\rho_3}-\frac{\rho_3}{2\rho_1},\quad\cos\alpha
_-=\frac{\rho_1\rho_2}{2\rho^2_3}-\frac{\rho_1}{2\rho_2}-\frac{\rho_2}{2
\rho_1},
\end{equation}
which are equivalent to the following condition in the complex plane:
\begin{equation}\rho^{-1}_1+\rho^{-1}_
2e^{i\alpha_-}+\rho^{-1}_3e^{-i\alpha_+}=0.\end{equation}
The CP-violating vacuum can only exist when the sides $\rho^{-1}_i$ satisfy triangle inequalities \cite{Branco:1980sz}.
 To determine which is the absolute minimum, we evaluate t%
he difference between \refeq{eq:VvevZ3} with the CP-conserving solution $V_{CPC}$ a%
nd the value associated with the CP-violating option $V_{CPV}$:

\begin{equation}V_{CPC}-V_{CPV}=\frac{(-\rho_1|\rho_2|-\rho_1|\rho_3|+\rho_2\rho_3)^2}{4\sigma_1\sigma_2\sigma_3},\end{equation}
As a result, the scalar potential of a $\jzc$ model generates spontaneous CPV whenever $\sigma_1\sigma_2\sigma_3>0$ and the $\rho^
{-1}_i$ satisfy triangle inequalities.

\subsection{$\jzs$ model -- explicit and spontaneous CPV\label{sSEC:CPscalar:Z2Z2}}We can write the %
scalar potential of a $\jzs$ model without loss of generality as
\begin{equation}\label{eq:potV:Z2Z2}
\begin{split}V(\phi)&=\mu^2_{ii}\jphi{i}{i}+\lambda_i\jphi{
i}{i}^2+\lambda_{ij}\jphi{i}{i}\jphi{j}{j}+\lambda'_{ij}\jphi{i}{j}\jphi{j%
}{i}\\&+\Big[\sigma_{12}\jphi{1}{2}^2+\sigma_{23}\jphi{2}{3}^2+\sigma_{31}
\jphi{3}{1}^2+h.c.\Big],
\end{split}
\end{equation}
with $\mu^2_{ii}$, $
\lambda_i$, $\lambda_{ij}$ and $\lambda'_{ij}$ real by hermiticity, $\sigma
_{12}$, $\sigma_{23}$ and $\sigma_{31}$ complex and implicit sums over $i<j
=1,2,3$.

By using the method described in the previous subsection, it can easily be 
checked that there is explicit CPV in this scalar potential provided that %
the product $\sigma_{12}\sigma_{23}\sigma_{31}$ is complex.

As before, once $\sigma_{12}\sigma_{23}\sigma_{31}$ is made real through th%
e imposition of CP as a symmetry of the full Lagrangian, we can always rep%
hase the scalar doublets to define $\sigma_{12}$, $\sigma_{23}$ and $\sigma
_{31}$ real. Thus, we can write the potential from which all vacuum phases 
are determined in the following way,
\begin{equation}\label{eq:VvevZ2Z2}
V(\braket{\phi})\supset
\frac{1}{2}\big[d_{12}\cos\big(\tilde{\alpha}_++\tilde{\alpha}_-\big)+d_{23
}\cos\tilde{\alpha}_++d_{31}\cos\tilde{\alpha}_-\big],
\end{equation}
where %
we introduced $\tilde{\alpha}_+=2(\alpha_3-\alpha_2)$, $\tilde{\alpha}_-=2(
\alpha_1-\alpha_3)$ and $d_{ij}=\sigma_{ij}v^2_iv^2_j$. Notice that after %
identifying $d_{ij}$, $\tilde{\alpha}_+$ and $\tilde{\alpha}_-$ with, resp%
ectively, $\rho_iv_1v_2v_3$, $\alpha_+$ and $\alpha_-$, this expression be%
comes identical to \refeq{eq:VvevZ3}. As such, the analysis performed for the $\jzc$
model in the last subsection is also valid for the $\jzs$ model. We conclu%
de by pointing out that, while distinct, the scalar sectors of both implem%
entations of the TP have identical CP properties, with the difference mani%
fest by the scalar masses and mixings\jf{The determination of these quanti%
ties from \refeq{eq:VvevZ3} and \refeq{eq:VvevZ2Z2} is beyond the scope of this paper.}.
We summarize our findings in the table below. 
\begin{table}[H]\centering\begin{tabular}{|c|cc|}\hline Symmetry&Exp%
licit CPV&Spontaneous CPV\\\hline$\jzc$&$\sigma_1\sigma_2\sigma_3$ complex&
$\sigma_1\sigma_2\sigma_3>0$ with triangle inequalities for $\rho^{-1}_i$\\
$\jzs$&$\sigma_{12}\sigma_{23}\sigma_{31}$ complex&$\sigma_{12}\sigma_{23}
\sigma_{31}>0$ with triangle inequalities for $d^{-1}_{ij}$\\\hline\end
{tabular}\caption{CP properties of the potentials given in \refeq{eq:potV:Z3} and %
\refeq{eq:potV:Z2Z2}. In the parameter space not covered above, there is no CPV.}\end{table}

\section{Complex CKM Matrix from Vacuum Phases\label{SEC:CKM}}
In a viable model of spontaneous CPV the vacuum phase(s) have to be able to generate a complex CKM matrix. This requirement stems from the fact that it has been shown that the CKM matrix has to be complex \cite{Botella:2005fc,Charles:2004jd,Bona:2005vz}, even if one allows for the presence of New physics beyond the SM. In this section we will prove that in the framework of models with a $\jzc$ or a $\jzs$ flavoured symmetry, one is able to generate a complex CKM matrix in agreement with experiment. We will also analyse carefully the relation between the generation of a complex CKM matrix and the appearance of SFCNC.\\  
Requiring CP invariance of the Yukawa lagrangian forces $\Gamma_i$ and $\Delta_i$ to be real. 
Starting with the $\jzc$ invariant model, it follows that the quark mass matrices have the form
\begin{equation}\label{eq:diagMf:SCPV}
\wMd=\mathrm{diag}\{e^{i\alpha_1},e^{i\alpha_2},e^{i\alpha_3}\}\rMd,\qquad 
\wMu=\mathrm{diag}\{e^{-i\alpha_1},e^{-i\alpha_3},e^{-i\alpha_2}\}\rMu,
\end{equation}
with $\rMd$, $\rMu$ real. Their bi-diagonalization (or polar decomposition) reads
\begin{equation}
O_{dL}^T\rMd O_{dR}=\mathrm{diag}\{m_d,m_s,m_b\}=\mMd,\quad
O_{uL}^T\rMd O_{uR}=\mathrm{diag}\{m_u,m_c,m_t\}=\mMu,
\end{equation}
with orthogonal matrices $O_X$, and thus
\begin{equation}\label{eq:UnitOrthTrans}
\begin{aligned}
&U_{dL}^\dagger\wMd O_{dR}=\mMd,\ \text{with}\ && U_{dL}=\mathrm{diag}\{e^{i\alpha_1},e^{i\alpha_2},e^{i\alpha_3}\}O_{dL},\\
&U_{uL}^\dagger\wMu O_{uR}=\mMu,\ \text{with}\ && U_{uL}=\mathrm{diag}\{e^{-i\alpha_1},e^{-i\alpha_3},e^{-i\alpha_2}\}O_{uL}.
\end{aligned}
\end{equation}
Consequently, the CKM matrix has the following form:
\begin{equation}\label{eq:CKM:Z3}
V=e^{i(\alpha_2+\alpha_3)}\,O^T_{uL}\,\mathrm{diag}\{e^{i(\alpha_++\alpha_-)},1,1\}\,O_{dL}.
\end{equation}
Apart from the irrelevant global phase $e^{i(\alpha_2+\alpha_3)}$, this structure was shown to be compatible with the current knowledge of the CKM matrix in \cite{Nebot:2018nqn}, including in particular the fact that the CKM matrix is irreducibly complex. The requirement that the CKM matrix is CP-violating places an important requirement on SFCNC: if SFCNC are absent in one sector, then the CKM matrix is necessarily CP conserving. In other words, tree level SFCNC in both up and down sectors are necessarily present in order to have a realistic CKM matrix.\\ 
We illustrate the reasoning behind this property with the down sector, the conclusion extends trivially to the up sector. Requiring flavour conservation in the down sector is equivalent to requiring that the projectors $P_j^{dL}$ which control SFCNC coincide with the canonical projectors $P_k$, not necessarily with $j=k$ (we recall that $[P_k]_{ab}=\delta_{ka}\delta_{kb}$): there is flavour conservation in the down sector if and only if $P_j^{dL}=P_{s(j)}$ with $s\in S_3$ a permutation. In that case, the elements of $O_{dL}$ are $[O_{dL}]_{jk}=\delta_{s(j)k}$ (the only non-vanishing elements are 1's, one per column and row, i.e. $O_{dL}$ represents the permutation $s$). In that case, in \refeq{eq:CKM:Z3}, $\mathrm{diag}\{e^{i(\alpha_++\alpha_-)},1,1\}\,O_{dL}=O_{dL}\mathrm{diag}\{s^{-1}(e^{i(\alpha_++\alpha_-)},1,1)\}$, i.e. $O_{dL}$ ``commutes'' with the diagonal phase matrix by permuting its elements, and thus all CP violation in \refeq{eq:CKM:Z3} can be rephased away, i.e. the CKM matrix is CP conserving. This property, i.e. that the absence of SFCNC in one quark sector is incompatible with a spontaneous origin of CP violation in the CKM matrix in this class of models, also appeared in the context of the 2HDM with spontaneous CP violation studied in \cite{Nebot:2018nqn}.

Regarding the $\jzs$ model, following the same arguments, one can obtain 
\begin{equation}
V=O^T_{uL}\mathrm{diag}\{e^{2i\alpha_1},e^{2i\alpha_2},e^{2i\alpha_3}\}O_{dL}.
\end{equation}
Notice that, besides an irrelevant global phase, in this model two relative vacuum phases remain in the intermediate diagonal matrix of phases. It is then clear that, as in the $\jzc$ invariant model, a realistic CKM matrix can be accommodated.\\ 
Concerning the new Yukawa couplings in section \ref{SEC:TP} and the complex orthonormal vectors $\nvec{i_1}{X}$ controlling them, it follows from \refeq{eq:UnitOrthTrans} that under the assumption of a spontaneous origin of CPV, the $\nvec{i_1}{X}$ reduce to \emph{real} orthonormal vectors (up to an irrelevant global phase). Consequently, rather than the 6 new real parameters that are required in the general case, only 3 new real parameters are necessary when CPV has a spontaneous origin (together, of course, with the vacuum expectation values, the CKM matrix and the fermion masses).

\section{Physical Implications\label{SEC:imp}}
In the previous sections the TP has been discussed and the possibility to combine it with a spontaneous origin of CPV analysed. The presence of (controlled) SFCNC and a common origin for CPV in the scalar and fermion sectors (necessarily present to obtain a realistic CKM matrix) have a wide range of interesting phenomenological consequences. Although these consequences exceed the scope of this paper, we nevertheless devote this section to a short analysis of some of them: in subsection \ref{sSEC:imp:mesons} we analyse sufficient conditions which guarantee that the SFCNC contributions to neutral meson mixings are in agreement with  phenomenological requirements; in subsection \ref{sSEC:imp:edm} we analyse how the contributions to light quark EDMs, in particular the ones involving flavour conserving Yukawa couplings, are sufficiently suppressed to respect neutron EDM constraints; finally, in subsection \ref{sSEC:BAU}, we discuss how CPV in these models can enhance the BAU with respect to SM expectations. For the discussions to follow, it is convenient to introduce the physical neutral scalars $h_j$ in the following manner,
\begin{equation}\label{eq:NeutralScalars}
\begin{pmatrix}h_0&h_1&h_2&h_3&h_4\end{pmatrix}^T=O
\begin{pmatrix}H^0&R&R'&I&I'\end{pmatrix}^T,
\end{equation}
where $O$ is a real $5\jx5$ orthogonal matrix %
and $h_0$ the scalar detected at the LHC \cite{Aad:2012tfa,Chatrchyan:2012%
xdj}. The mixing between CP-even and odd states shown above is implied by %
the potentials of these models. 
From \refeq{eq:YukNeutral:0}, the physical Yukawa couplings read
\begin{equation}\label{eq:YukNeutral:1}
\mathscr{L}_{h_a\bar qq}=-h_a\sum_{f=u,d} \bar f_{L}Y^a_f f_{R}+\mathrm{h.c.}\,,
\end{equation}
where
\begin{equation}\label{eq:YukNeutral:2}
Y^a_f=O_{a0}\mM{f}/v+(O_{a1}+i\epsilon_fO_{a3})N_f/v'+(O_{a2}+i\epsilon_fO_{a4})N'_f/v''\,,
\end{equation}
and, again, $\epsilon_d=-\epsilon_u=1$. Equations \eqref{eq:Nq:Z3} and \eqref{eq:Nq:Z2Z2} can be condensed into
\begin{multline}\label{eq:YukNeutral:3}
(Y^a_f)_{jk}=\frac{m_{f_k}}{v}
\bigg\{O_{a0}\delta_{jk}+(O_{a1}+i\epsilon_fO_{a3})\left[\frac{v_2}{v_1}\big(P^{fL}_1\big)_{
jk}-\frac{v_1}{v_2}\big(P^{fL}_{\alpha}\big)_{jk}\right]\frac{v}{v'}\\
+(O_{a2}+i\epsilon_fO_{a4})\left[\delta_{jk}-\frac{v^2}{v_3^2}\big(P^{fL}_{5-\alpha}\big)_{jk}\right]\frac{v}{v''}\bigg\}\,,
\end{multline}
where $\alpha=3$ when $f=u$ in the $\jzc$ symmetric model, and $\alpha=2$ otherwise.

\subsection{SFCNC -- mixing in meson-antimeson systems\label{sSEC:imp:mesons}}We will probe the s%
trength of the tree-level SFCNC in the $\jzc$ and $\jzs$ models via an eva%
luation of the new contributions to the amplitude of the mixing in meson-a%
ntimeson systems, with a comprehensive analysis beyond the scope of this p%
aper. From studying \refeq{eq:YukNeutral:3}, we conclude that these models are suppressed 
by the following factors, before performing any calculation:\begin{itemize}
\item The mass $m_j$ leading to proportionality to the largest Yukawa in t%
he system;\item The term $O_{ab}\pm iO_{a,b+2}$ with magnitude below $1$ s%
ince $O$ is real and orthogonal;\item A factor of $(P^{fL}_i)_{jk}=(U_{fL})
^*_{ij}(U_{fL})_{ik}$ that ranges from $0$ to $1/2$ in absolute value;\item
Ratios of vacuum expectation values that perturbative unitarity may constr%
ain.\end{itemize}In appendix \ref{APP:constraints}, we show that all $\jzc$ and $\jzs$ models %
which satisfy the relation below respect the current experimental bounds r%
elated to meson-antimeson systems\jf{This result was obtained from the $D^0
-\bar{D}^0$ system that turned out to be the most stringent.},
\begin{equation}\left(\frac{2v^2}{v^2_1}+\frac{2v^2}{v^2_2}+\frac{v^2}{v^2_
3}-\frac{3v^2}{v'^2}\right)\sum^4_{a=0}\frac{1-O^2_{a0}}{x^2_a}<2\jx10^{-4}
,\end{equation}with $x_a=m_{h_a}/m_{h_0}$ controlled by the absence of a d%
ecoupling limit in both models (see appendix \ref{APP:scalarmasses}). Despite not providing an %
exclusion region, the expression above remains interesting as it illustrat%
es the degree of cancellations required by these models, be it in their fl%
avour sector, scalar potential or a combination of the two.

\subsection{CPV -- electric dipole moments\label{sSEC:imp:edm}}
In the context of 2HDM, one and two loop contributions to the EDM of light quarks can be excessively large\footnote{Since we do not consider the lepton sector in this work, contributions to the EDM of the electron are not discussed.}. Having Yukawa couplings proportional to the fermion masses is typically sufficient to obtain one loop contributions adequately suppressed. This suppression can be partially circumvented in so-called Barr-Zee two loop contributions \cite{Barr:1990vd,Chang:1990sf,Cheung:2009fc}, in particular the dominant ones which involve flavour conserving couplings and neutral scalars\footnote{Similar contributions with flavour changing couplings of either neutral or charged scalars are further suppressed.}. For generic flavour conserving Yukawa couplings of the form
\begin{equation}
\mathscr L=-S\bar f(A_f^S+iB_f^S\gamma_5)f\,,
\end{equation}
with $S$ a neutral scalar, the contribution to the EDM $d_q$ of the light quark $q$ is
\begin{equation}\label{eq:EDM:0}
d_q^S=-\frac{\alpha^2}{8\pi^2s_W^2}\frac{v^2}{M_W^2}\sum_f N_c^fQ_f^2\frac{1}{m_f}
\left\{
B_q^S A_f^S F(z_{fS})+A_q^S B_f^S G(z_{fS})
\right\},
\end{equation}
with $N_c^f$ and $Q_f$ the number of colours and the electric charge of the virtual fermion $f$, $z_{fS}=m_f^2/m_S^2$, and $F$ and $G$, the loop functions\footnote{For scalar masses $m_S\in[0.2;1.0]$ TeV, we have $F(z_{tS}),G(z_{tS})\in[0.1;1.0]$ while $F(z_{bS}),G(z_{bS})\in[5\times 10^{-4};10^{-2}]$.}. With the Yukawa couplings in \refeq{eq:YukNeutral:2}, we have
\begin{equation}
\begin{aligned}
&A_{f_j}^{h_a}=\left\{O_{a0}\frac{m_{f_j}}{v}+O_{a1}\frac{(N_f)_{jj}}{v'}+O_{a2}\frac{(N_f')_{jj}}{v''}\right\},\\
&B_{f_j}^{h_a}=\epsilon_f\left\{O_{a3}\frac{(N_f)_{jj}}{v'}+O_{a4}\frac{(N_f')_{jj}}{v''}\right\}.
\end{aligned}
\end{equation}
It is then clear that in the products of scalar $\times$ pseudoscalar couplings in \refeq{eq:EDM:0}, there is one light fermion mass suppression factor, and another relevant suppression to be noticed: the products of matrix elements $O_{jk}$ necessarily involve one element which mixes the fields $\{H^0,R,R^{'}\}$ and $\{I,I^{'}\}$ to give the mass eigenstates in \refeq{eq:NeutralScalars} (for a CP conserving scalar sector, these would be, respectively, CP-even and CP-odd fields, and would not mix). Furthermore, contributions from the different scalars can easily interfere destructively or cancel. In conclusion, EDMs of light quarks are not a source of concern for the viability of the models under consideration, even though they can have some impact in excluding regions of parameter space.

\subsection{Estimation of the Enhancement of the BAU\label{sSEC:BAU}}
It has been established that in the SM one cannot obtain a BAU sufficient to be in agreement with the value derived from the CMB measurements. The reason for this shortcoming of the SM has to do with the following:
\begin{enumerate}
\item[i)] CP violation in the SM is too small.
\item[ii)] The electroweak phase transition is not strongly first order, as required by electroweak baryogenesis.
\end{enumerate}
In this paper we will not analyse ii) and concentrate on i), where we point out the importance of new sources of CP violation which arise in TP models. It has been shown that in the SM, for an arbitrary number of generations, a necessary condition to have CP invariance is that the following WB invariant vanishes \cite{Bernabeu:1986fc}:
\begin{equation}
I_{\rm SM}=\text{Tr}\left[\wMd\wMdag{d},\wMu\wMdag{u}\right]^3\,.
\end{equation}
For three generations \cite{Jarlskog:1985ht} the above condition becomes a necessary and sufficient condition for CP invariance. In the quark mass eigenstate bases, one obtains
\begin{equation}\label{eq:InvSM}
\mathrm{Tr}\left[\wMd\wMdag{d},\wMu\wMdag{u}\right]^3=6i\,\Delta m_{tc}\Delta m_{tu}\Delta m_{cu}\Delta m_{bs}\Delta m_{bd}\Delta m_{sd}\,\text{Im}Q,
\end{equation}
with $\Delta m_{jk}=m^2_j-m^2_k$ and $Q$ denoting a rephasing invariant quartet of the CKM matrix $V$ \cite{Branco:1999fs}. Noting that $I_{\rm SM}$ has dimensions (Mass)$^{12}$, it is plausible that
\begin{equation}\label{eq:BAUSM}
[\mathrm{BAU}]_{\rm SM}\propto \frac{I_{\rm SM}}{v^{12}}\sim 10^{-19}.
\end{equation}
The smallness of CP violation in the SM has to do with the smallness of quark masses compared with the electroweak breaking scale. In the TP models one can construct CP odd WB invariants of a much lower dimension, such as:
\begin{equation}\label{eq:BAUest:0}
\mathrm{Im}\left[\mathrm{Tr}\left(N'^0_d\wMdag{d}\wMu\wMdag{u}\right)\right]=
-\frac{v^2}{v^2_3}\mathrm{Im}\left[\mathrm{Tr}\left(P^{dL}_3\mMd^2V^\dagger \mMu^2V\right)\right].
\end{equation}
It is straightforward to check that this expression is WB invariant by using the following transformation laws,
\begin{equation}
N'^0_d\to W^\dagger_LN'^0_dW_{dR},\quad \wM{f}\to W^\dagger_L\wM{f}W_{fR}.
\end{equation}
We proceed by keeping the leading term in the fermion masses and in the Wolfenstein parameter $\lambda$ \cite{Wolfenstein:1983yz} to find\jf{Since $N'^0_d$
is given by the same expression in the two models, the argument will be valid for both.}
\begin{equation}\label{eq:BAUest:1}
\mathrm{Im}\left[\mathrm{Tr}\left(N'^0_d\wMdag{d}\wMu\wMdag{u}\right)\right]=
-m^2_bm^2_t\frac{v^2}{v^2_3}\,\mathrm{Im}\left[(U_{dL})^*_{32}(U_{dL})_{33}V_{ts}V^*_{tb}\right].
\end{equation}
Using equations \eqref{eq:InvSM}, \eqref{eq:BAUSM}, \eqref{eq:BAUest:1} we obtain the following enhancement factor
\begin{equation}
\frac{\mathrm{BAU}_\mathrm{TP}}{\mathrm{BAU}_\mathrm{SM}}\propto 
10^{15}\frac{v^2}{v^2_3}\left|(U_{dL})^*_{32}(U_{dL})_{33}\right|\,\left|\sin\mathrm{arg}\left[(U_{dL})^*_{32}(U_{dL})_{33}V_{ts}V^*_{tb}\right]
\right|.
\end{equation}
This enhancement suggests that one may find an adequate size of the BAU in TP models, once the problem of having a correct electroweak phase transition is solved.

\section{Conclusions}
We conjectured that there is an analogy between scalars and fermions which makes it likely that there are also three scalar doublets. A Trinity Principle is introduced which, given the fact that there are no massless quarks, requires the existence of a minimum of three Higgs doublets. This principle is similar to the Glashow and Weinberg proposal of NFC in the scalar sector of a 2HDM, with the crucial difference that the TP requires flavoured symmetries to be implemented. In the minimal realisation, the same set of scalar doublets couples to both up and down quarks. Another possibility is having three scalar doublets coupling to the up quarks and another set of three scalar doublets coupling to down quarks. This realisation of the TP can be incorporated into a supersymmetric extension of the SM. 

We give two explicit examples of models with three Higgs doublets which satisfy the TP. The first is invariant under a $\jzc$ flavoured symmetry, and is related to jBGL models. The second employs a $\jzs$ symmetry, being connected to 3BGL models. In both, one can have either explicit or spontaneous CPV in the scalar sector, with the vacuum associated to the latter possibility, capable of generating a complex CKM matrix. These models are the minimal extension of the SM with this feature of generating spontaneous CP violation and a complex CKM, without introducing in the Lagrangian soft breaking terms. We have studied in detail the deep connection between the generation of a complex CKM from vacuum phases and the appearance of tree level SFCNC.

We have shown that there are new sources of CP violation which lead to a significant enhancement of the BAU in both models. We also identify several suppression factors which control the strength of their tree-level SFCNC, rendering them plausible extensions of the SM.

\section*{Acknowledgments}

JA, GCB and MN acknowledge support from Funda\c{c}\~ao para a Ci\^encia e a Tecnologia (FCT, Portugal) through the projects CFTP-FCT Unit 777 (UID/FIS/00777/2013, UID/FIS/00777/2019), CERN/FIS-PAR/0004/2017 and PTDC/FIS-PAR/29436/2017 which are partially funded through POCTI (FEDER), COMPETE, QREN and EU. The work of JA is funded through the doctoral FCT grant SFRH/BD/139937/2018.
FJB and MN acknowledge support from Spanish grant FPA2017-85140-C3-3-P (AEI/FEDER, UE) and PROMETEO 2019-113 (Generalitat Valenciana).

\clearpage\appendix
\section{Defining Implementations of the TP\label{APP:implem}}

\subsection{Symmetries\label{sAPP:implem:sym}}
As discussed in section \ref{SEC:TP}, one can a priori conceive different implementations of the TP depending on the number of doublets $\phi_j$ which couple to the same quark doublet $q_{Lk}$ in both up and down quark sectors. In terms of the Yukawa coupling matrices, this simply corresponds to both $\Gamma_j$ and $\Delta_j$ having the same non-vanishing row $k$. The implementations of the TP correspond to the different permutations $p\in S_3$ such that the Yukawa matrices in the up quark sector $\Delta_j$ read
\begin{equation}\label{eq:TPcases:matrices}
\Delta_{p(1)}\sim\Gamma_1,\ \Delta_{p(2)}\sim\Gamma_2,\ \Delta_{p(3)}\sim\Gamma_3,
\end{equation}
with the understanding that $\Delta_{p(j)}\sim\Gamma_j$ means that both matrices have the same vanishing elements, i.e. the same texture. The different implementations of the TP correspond to three different cases\footnote{The notation corresponds to the standard decomposition in cycles, e.g. $p=(123)$ stands for $p(1)=2$, $p(2)=3$, $p(3)=1$, etc.}.
\begin{enumerate}
\item For $p=(1)(2)(3)$ (that is $p=e$, the identity), all scalar doublets have a Yukawa matrix with the same texture in both sectors.
\item For $p=(3)(12)$, $(2)(13)$, $(1)(23)$, only one scalar doublet has a Yukawa matrix with the same texture in both sectors.
\item For $p=(123)$, $(132)$, no scalar doublet has a Yukawa matrix with the same texture in both sectors.
\end{enumerate}
Without loss of generality one can redefine the labels of the scalar and quark doublets such that: (i) the Yukawa matrices of down quarks read
\begin{equation}\label{eq:downmatrices}
\Gamma_1\sim\begin{pmatrix}\jx&\jx&\jx\\ 0&0&0\\ 0&0&0\end{pmatrix},\ 
\Gamma_2\sim\begin{pmatrix}0&0&0\\ \jx&\jx&\jx\\ 0&0&0\end{pmatrix},\ 
\Gamma_3\sim\begin{pmatrix}0&0&0\\ 0&0&0\\ \jx&\jx&\jx\end{pmatrix},
\end{equation}
and (ii) only one instance within each case needs to be considered. Consequently, the different possible implementations of the TP that we consider are just $p=e$ for case 1, $p=(1)(23)$ for case 2, and $p=(123)$ for case 3.\\ 
The question now is, how can one obtain these implementations through symmetries?\\ 
According to the TP, the candidate symmetry cannot differentiate among the different generations of right-handed fields in each sector. It should allow or forbid rows in each Yukawa matrix, not relate different rows in the same or different matrices.\\ 
We start by requiring invariance under transformations of the form
\begin{equation}\label{eq:sym:Zn}
\phi_a\mapsto \omega^{\varphi_a}\phi_a,\quad q_{Lb}\mapsto\omega^{Q_b}q_{Lb},\quad d_{R\alpha}\mapsto\omega^{D}d_{R\alpha},\quad u_{R\beta}\mapsto\omega^{U}u_{R\beta},
\end{equation}
where $\omega=e^{i\frac{2\pi}{n}}$ for some $n\in\mathbb{N}$. That is, we focus on $\mathbb{Z}_n$ transformations, for which $\varphi_a$, $Q_b$, $U$ and $D$ are the corresponding ``charges''. Although one could have considered continuous $U(1)$ transformations (with $\omega=e^{i\tau}$, $\tau\in\mathbb{R}$), in order to avoid massless scalars, and attending to the possibility of having a spontaneous origin of CPV, we only consider discrete transformations. Equation \eqref{eq:sym:Zn} is an appropriate starting point to analyse how the TP can be implemented in terms of symmetries because (i) an abelian symmetry of the Yukawa sector of a multi-Higgs doublets model can always be brought to a form similar to \refeq{eq:sym:Zn} \cite{Ferreira:2010ir,Serodio:2013gka} (with respect to that general case, we concentrate on finite rather than continuous symmetries, and in addition impose no generation dependence on the charges of the right-handed fields), and (ii) any finite abelian group is isomorphic to the direct product of finitely many cyclic groups (that is $\mathbb{Z}_n$ factors). Furthermore, for a finite non-abelian symmetry group, $\mathbb{Z}_n$ subgroups are necessarily present. It is to be noticed that if the candidate symmetry consists of $\mathbb{Z}_n$ alone, the TP requires that all $\varphi_a$ are different and also that all $Q_b$ are different. If instead the candidate symmetry is the (direct) product of several $\mathbb{Z}_n$ factors (not necessarily with the same $n$), the previous requirement is relaxed: in that case the minimal requirement to implement the TP is either that all $\varphi_a$ are not equal or that all $Q_b$ are not equal. What are the consequences of requiring invariance under \refeq{eq:sym:Zn}? As a first step, we concentrate on the conditions for the allowed rows of the different Yukawa matrices\footnote{These are not \emph{all} the conditions but, with the benefit of hindsight, they are sufficient to orient the search of symmetries that do implement the TP in cases 1 and 2, and they are also sufficient to show that case 3 cannot be implemented through a symmetry.}. For the down Yukawa matrices in \refeq{eq:downmatrices}, they read \footnote{Notice that these relations among ``charges'' should be understood as being ``modulus $n$'', since in \refeq{eq:sym:Zn}, $\omega^n=1$.}
\begin{equation}\label{eq:symcond:down}
\varphi_1-Q_1+D=0,\ \varphi_2-Q_2+D=0,\ \varphi_3-Q_3+D=0,
\end{equation}
while for the up Yukawa matrices, following \refeq{eq:TPcases:matrices}, we have
\begin{equation}\label{eq:symcond:up}
-\varphi_{p(1)}-Q_1+U=0,\ -\varphi_{p(2)}-Q_2+U=0,\ -\varphi_{p(3)}-Q_3+U=0.
\end{equation}
If follows that
\begin{equation}\label{eq:symcond:gen:1}
\varphi_1+\varphi_{p(1)}=\varphi_2+\varphi_{p(2)}=\varphi_3+\varphi_{p(3)}\ ,
\end{equation}
and
\begin{equation}\label{eq:symcond:gen:2}
2Q_1-\varphi_1+\varphi_{p(1)}=2Q_2-\varphi_2+\varphi_{p(2)}=2Q_3-\varphi_3+\varphi_{p(3)}\ .
\end{equation}
At this point one has to consider separately the different cases.
\begin{enumerate}
\item For case 1, $p=(1)(2)(3)$, and \refeqs{eq:symcond:gen:1}{eq:symcond:gen:2} give
\begin{equation}
2\varphi_1=2\varphi_2=2\varphi_3,\quad \text{and}\quad 2Q_1=2Q_2=2Q_3\ .
\end{equation}
It is clear that a single $\mathbb{Z}_n$ factor is not sufficient for this implementation of the TP; furthermore, to fulfill the previous conditions non-trivially, i.e. with not all charges equal, $n$ has to be, necessarily, even. The minimal realization through a $\mathbb{Z}_2\times\mathbb{Z}_2$ symmetry is the model introduced in section \ref{sSEC:TP:Z2Z2flavour}.
\item For case 2, $p=(1)(23)$ and \refeqs{eq:symcond:gen:1}{eq:symcond:gen:2} give
\begin{equation}
2\varphi_1=\varphi_2+\varphi_3,\quad \text{and}\quad 2Q_1=2Q_2-\varphi_2+\varphi_{3}=2Q_3-\varphi_3+\varphi_{2}\ .
\end{equation}
In this case, a single $\mathbb{Z}_n$ is sufficient: one can easily check that a simple charge assignment such as $\varphi_1=Q_1=U=D=0$, $\varphi_2=Q_2=1$ and $\varphi_3=Q_3=-1$ satisfies the previous conditions and forces all elements other than the ones in the selected rows (of the corresponding matrices) to vanish, thus giving this second implementation of the TP, addressed in section \ref{sSEC:TP:Z3flavour}. 
\item For case 3, $p=(123)$, \refeq{eq:symcond:gen:1} gives
\begin{equation}\label{eq:case3:phi}
\varphi_1=\varphi_2=\varphi_3,
\end{equation}
and thus, from \refeq{eq:symcond:down}, 
\begin{equation}\label{eq:case3:Q}
Q_1=Q_2=Q_3.
\end{equation}
The previous conditions have been obtained by considering only the fact that, within this implementation of the TP, (i) different rows of the up and down Yukawa matrices of a given scalar doublet must be allowed, and (ii) different rows must be allowed when the couplings of a $q_{Lj}$ with the different $\phi_{k}$ are considered in each sector. The implication of \refeqs{eq:case3:phi}{eq:case3:Q} is then clear: these requirements are only satisfied in the most trivial manner, when all rows are equally allowed in all Yukawa matrices. It follows that this implementation of the TP cannot be obtained from a symmetry (as argued in precedence, considering the $\mathbb{Z}_n$ transformation in \refeq{eq:sym:Zn} is then sufficient to discard both abelian and non-abelian finite symmetries): it is for this reason that this implementation of the TP is not addressed.
\end{enumerate}

\subsection{Weak basis invariant conditions\label{sAPP:implem:wb}}
As anticipated in section \ref{SEC:TP}, \refeq{eq:WB:Z3} and \refeq{eq:WB:Z2Z2} provide WB independent definitions, respectively, of the $\jzc$ and $\jzs$ invariant models: this section gives the corresponding proof. Since the arguments for both models
are similar, we can focus on the latter. As such, we must show that it is %
both sufficient and necessary for a given set of textures to satisfy %
\refeq{eq:WB:Z2Z2} for it to belong to a $\jzs$ model.

Given that, by definition, a $\jzs$ model must have a WB defined by the Yu%
kawa couplings of \refeq{eq:Yuk:Z2Z2}, we can show the necessity of \refeq{eq:WB:Z2Z2} through %
its direct evaluation.

To prove the sufficiency of \refeq{eq:WB:Z2Z2}, we start by using the polar decompos%
ition\begin{equation}\Gamma_i=U^d_iD^d_iW^d_i,\quad\Delta_i=U^u_iD^u_iW^u_i
,\end{equation}where $D^f_i$ is diagonal and the rest are unitary. Then, we
evaluate the commutators\begin{equation}\big[\Gamma_i\Gamma^\dagger_i,
\Gamma_j\Gamma^\dagger_j\big]=\big[\Gamma_i\Gamma^\dagger_i,\Delta_j\Delta^
\dagger_j\big]=0,\quad i\neq j,\end{equation}to determine that $U^d_i=U^u_i
=U$. As such, we can rewrite \refeq{eq:WB:Z2Z2} in diagonal form,\begin{equation}D^d_
iD^d_j=D^d_iD^u_j=0,\quad\quad i\neq j.\end{equation}It is clear that, up %
to a non-physical permutation of the scalars, the only solution with three 
massive up-type quarks is given by\begin{equation}D^d_1\propto D^u_1\propto
\mathrm{diag}\{\jx,0,0\},\quad D^d_2\propto D^u_2\propto\mathrm{diag}\{0,
\jx,0\},\quad D^d_3\propto D^u_3\propto\mathrm{diag}\{0,0,\jx\}.
\end{equation}Thus, we can write the Yukawa couplings of any 3HDM which sa%
tisfies \refeq{eq:WB:Z2Z2} as\begin{equation}\begin{split}\Gamma_1=U\begin{pmatrix}
\jx&\jx&\jx\\0&0&0\\0&0&0\end{pmatrix},\quad\,\Gamma_2&=U\begin{pmatrix}0&0
&0\\\jx&\jx&\jx\\0&0&0\end{pmatrix},\quad\,\Gamma_3=U\begin{pmatrix}0&0&0\\
0&0&0\\\jx&\jx&\jx\end{pmatrix},\\\Delta_1=U\begin{pmatrix}\jx&\jx&\jx\\0&0
&0\\0&0&0\end{pmatrix},\quad\Delta_2&=U\begin{pmatrix}0&0&0\\\jx&\jx&\jx\\0
&0&0\end{pmatrix},\quad\Delta_3=U\begin{pmatrix}0&0&0\\0&0&0\\\jx&\jx&\jx
\end{pmatrix}.\end{split}\end{equation}We finish this demonstration by not%
ing that \refeq{eq:Yuk:Z2Z2} can be recovered after performing the WB transformation $
q^0_L\rightarrow Uq^0_L$.

\subsection{RGE stability of TP implementations\label{sAPP:implem:RGE}}
To close this section on the implementations of the TP, we analyse their stability under one loop renormalization group evolution (RGE). The RGE equations read
\begin{align}
\label{eq:RGE:down}
&\mathcal D\Gamma_i=a_\Gamma\Gamma_i+\sum_{j=1}^3\left[\alpha_{ij}\Gamma_j-2\Delta_j\Delta_i^\dagger\Gamma_j+\Gamma_i\Gamma_j^\dagger\Gamma_j+\frac{1}{2}\Delta_j\Delta_j^\dagger\Gamma_i+\frac{1}{2}\Gamma_j\Gamma_j^\dagger\Gamma_i\right],\\
\label{eq:RGE:up}
&\mathcal D\Delta_i=a_\Delta\Delta_i+\sum_{j=1}^3\left[\alpha_{ji}\Delta_j-2\Gamma_j\Gamma_i^\dagger\Delta_j+\Delta_i\Delta_j^\dagger\Delta_j+\frac{1}{2}\Gamma_j\Gamma_j^\dagger\Delta_i+\frac{1}{2}\Delta_j\Delta_j^\dagger\Delta_i\right],
\end{align}
with $\mathcal D\equiv 16\pi^2\frac{d}{d\ln\mu}$, $a_\Gamma$ and $a_\Delta$ constants, and $\alpha_{ij}\equiv\text{Tr}(\Gamma_i\Gamma_j^\dagger+\Delta_i^\dagger\Delta_j)$. Attending to the initial discussion in subappendix \ref{sAPP:implem:sym}, an implementation of the TP is defined by
\begin{equation}\label{eq:TP:impl:RGE}
\Gamma_i=P_i\Gamma_i\quad\text{and}\quad \Delta_i=P_{s(i)}\Delta_i,\quad i=1,2,3,
\end{equation}
with a permutation $s\in S_3$, and $P_i$ and $P_{s(i)}$ canonical projectors\footnote{It is useful to recall here that $P_i^\dagger=P_i$ and $P_jP_k=\delta_{jk}P_k$ (no sum).}. Then, stability under RGE means
\begin{equation}
\mathcal D\Gamma_i=P_i\,X_{(i)}\quad\text{and}\quad \mathcal D\Delta_i=P_{s(i)}\,Y_{(i)},\quad i=1,2,3,
\end{equation}
with some matrices $X_{(i)}$ and $Y_{(i)}$. The relevant point is, of course, that the corrections to each Yukawa matrices due to the RGE also respect the same TP requirement. We already know that two TP implementations can be obtained from a symmetry and thus we expect them to be stable, while the third cannot be obtained from a symmetry, and we do not expect it to be stable. Using \refeq{eq:TP:impl:RGE} and the properties of the projectors $P_i$, one can show that
\begin{equation}
\mathcal D\Gamma_i=\Gamma_i\bigg(a_\Gamma+3\alpha_{ii}+\frac{1}{2}\Gamma_i^\dagger\Gamma_i+\sum_{j=1}^3\Gamma_j^\dagger\Gamma_j\bigg)+\frac{1}{2}P_i\Delta_{s^{-1}(i)}\Delta_{s^{-1}(i)}^\dagger\Gamma_i-2P_{s^2(i)}\Delta_{s(i)}\Delta_i^\dagger\Gamma_{s(i)},
\end{equation}
and
\begin{equation}
\mathcal D\Delta_i=\Delta_i\bigg(a_\Delta+3\alpha_{ii}+\frac{1}{2}\Delta_i^\dagger\Delta_i+\sum_{j=1}^3\Delta_j^\dagger\Delta_j\bigg)+\frac{1}{2}P_{s(i)}\Gamma_{s(i)}\Gamma_{s(i)}^\dagger\Delta_i-2P_{s^{-1}(i)}\Gamma_i\Gamma_i^\dagger\Delta_{s^{-1}(i)}.
\end{equation}
Every term except the last in each of the previous equations respects the stability requirement under RGE. For these last terms, the question of stability is reduced to having $s^2(i)=i$ or equivalently $s^{-1}(i)=s(i)$ for $i=1,2,3$. It follows that the TP implementations corresponding to $s=(1)(2)(3)$ (the $\jzs$ symmetric case) and $s=(1)(23)$ or $(2)(13)$ or $(3)(12)$ (the $\jzc$ symmetric case) are trivially stable, as expected. Furthermore, the last TP implementation, which cannot be obtained from a symmetry and which corresponds to $s=(123)$ or $s=(132)$, is not stable since $s(i)\neq s^{-1}(i)$.

\section{Deriving Experimental Constraints\label{APP:constraints}}
We start by using \refeq{eq:NeutralScalars} and \refeq{eq:YukNeutral:1} to construct the effective Hamiltonian
\begin{equation}\label{eq:Heff}
\mathcal{H}_{eff}=\sum^4_{a=0}\frac{\big(Y^{f\dagger}_{a}\big)^2_{ji}}{m^2_{h_a}}\big(\bar{j}\gamma_Li\big)^2,
\end{equation}
in which we considered the meson $P^
0=f_i\bar{f}_j$ while neglecting $m_i$ in regards to $m_j$. By requiring t%
he new tree-level contributions to be inferior to the amplitude of mixing, 
we obtain the constraint
\begin{equation}\label{eq:MesonMix:ineq}
\left|\sum^4_{a=0}\frac{\big(Y_{a}^{f\dagger}\big)^2_{ji}}{m^2_{h_a}}\right|<\frac{12m^2_j\Delta m_P}{5m^3_Pf^2_P}.
\end{equation}
While an exact expression could be found for the left%
-handed side of the relation above, it turns out to be more useful to cont%
rol it in the following manner,
\begin{multline}
\left|\sum^4_{a=0}\frac{\big(Y^{f\dagger}_{a}\big)^2_{ji}}{m^2_{h_a}}\right|\leq
\sum^4_{a=0}\frac{|Y^f_{a}|^2_{ij}}{m^2_{h_a}}\leq
\sum^4_{a=0}\frac{m^2_j}{v'^2m^2_{h_a}}\big(|O_{a1}-i\epsilon_fO_{a3}|^2+|O_{a2}-i\epsilon_fO_{a4}|^2
\big)\jx\\
\bigg[\Big(\frac{v'^4}{v^2_1v^2_2}-2\Big)\big(|(P^{fL}_1)_{ij}|^2+|(P^{fL}_q)_{ij}|^2\big)+\frac{v^2}{v^2_3}|(P^{fL}_{5-q})_{ij}|^2\bigg]  \\
\leq \frac{m^2_j}{4v^2}\left(\frac{2}{v^2_1}+\frac{2}{v^2_2}+\frac{1}{v^2_3}-\frac{3}{v'^2}\right)\sum^4_{a=0}\frac{1-O^2_{a0}}{m^2_{h_a}}.
\end{multline}
In the derivation of this result, we started by using the tr%
iangle inequality, followed by the Cauchy-Schwarz inequality, and finally %
the orthonormality relations $\sum^4_{b=0}O^2_{ab}=1$ and $|(P^X_i)_{jk}|\leq
1/2$. As such, we conclude that whenever a $\jzc$ or $\jzs$ model satisfies
\begin{equation}\left(\frac{2v^2}{v^2_1}+\frac{2v^2}{v^2_2}+\frac{v^2}{v^2_
3}-\frac{3v^2}{v'^2}\right)\sum^4_{a=0}\frac{1-O^2_{a0}}{x^2_a}<\frac{48v^2
m^2_h\Delta m_P}{5m^3_Pf^2_P},\end{equation}it will never saturate \refeq{eq:MesonMix:ineq}\jf{Notice that $2v^2/v^2_1+2v^2/v^2_2+v^2/v^2_3-3v^2/v'^2\geq2(3+\sqrt{5}
)$.}.

\section{Controlling the Scalar Masses\label{APP:scalarmasses}}
By expanding the degrees of freedom 
in the scalar fields of \refeq{eq:potV:Z3} and \refeq{eq:potV:Z2Z2},\begin{equation}\phi_i=e^{i
\alpha_i}\begin{pmatrix}\varphi^+_i\\\frac{1}{\sqrt{2}}(v_i+\tau_i+i\eta_i)
\end{pmatrix},\end{equation}we can write the mass terms for the neutral sc%
alars as\begin{equation}\mathscr{L}_{mass}=-\frac{1}{2}\big(M^2_\tau\big)_{
ij}\tau_i\tau_j-\frac{1}{2}\big(M^2_\eta\big)_{ij}\eta_i\eta_j-\big(M^2_{
\tau\eta}\big)_{ij}\tau_i\eta_j.\end{equation}After using the minimization 
conditions of \refeq{eq:potV:Z3} with respect to $v_i$,\begin{equation}\begin{split}
\mu^2_{11}&=-\lambda_1v^2_1-\lambda''_{12}\frac{v_2}{2v_1}-\lambda''_{13}
\frac{v_3}{2v_1}-\big(2\Upsilon_1+\Upsilon_2+\Upsilon_3\big)\frac{v_2v_3}{2
v_1},\\\mu^2_{22}&=-\lambda_2v^2_2-\lambda''_{12}\frac{v_1}{2v_2}-\lambda''
_{23}\frac{v_3}{2v_2}-\big(\Upsilon_1+2\Upsilon_2+\Upsilon_3\big)\frac{v_3v
_1}{2v_2},\\\mu^2_{33}&=-\lambda_3v^2_3-\lambda''_{13}\frac{v_1}{2v_3}-
\lambda''_{23}\frac{v_2}{2v_3}-\big(\Upsilon_1+\Upsilon_2+2\Upsilon_3\big)
\frac{v_1v_2}{2v_3},\end{split}\end{equation}where $\Upsilon_1=\rho_1\cos(
\alpha_++\alpha_-)$, $\Upsilon_2=\rho_2\cos\alpha_+$ and $\Upsilon_3=\rho_3
\cos\alpha_-$, and those for \refeq{eq:potV:Z2Z2},\begin{equation}\begin{split}\mu^2_{1
1}&=-\lambda_1v^2_1-\big(\lambda''_{12}+2\Upsilon_{12}\big)\frac{v_2}{2v_1}
-\big(\lambda''_{13}+2\Upsilon_{31}\big)\frac{v_3}{2v_1},\\\mu^2_{22}&=-
\lambda_2v^2_2-\big(\lambda''_{13}+2\Upsilon_{12}\big)\frac{v_1}{2v_2}-\big
(\lambda''_{23}+2\Upsilon_{23}\big)\frac{v_3}{2v_2},\\\mu^2_{33}&=-\lambda_
3v^2_3-\big(\lambda''_{13}+2\Upsilon_{31}\big)\frac{v_1}{2v_3}-\big(\lambda
''_{23}+2\Upsilon_{23}\big)\frac{v_2}{2v_3},\end{split}\end{equation}in wh%
ich $\Upsilon_{12}=\sigma_{12}v_1v_2\cos(\tilde{\alpha}_++\tilde{\alpha}_-)
$, $\Upsilon_{23}=\sigma_{23}v_2v_3\cos\tilde{\alpha}_+$ and $\Upsilon_{31}
=\sigma_{31}v_3v_1\cos\tilde{\alpha}_-$.\\ 
We can evaluate the sum of all neutral scalar masses in the $\jz_3$ model as the following trace (the massless pseudo-Goldstone boson $G^0$ is a linear combination of the $\eta_i$'s),
\begin{equation}\mathrm{tr}\big(M^2_\tau\big)+\mathrm{tr}\big(M^2_\eta
\big)=2(\lambda_1v^2_1+\lambda_2v^2_2+\lambda_3v^2_3)+\frac{\sigma^2_1
\sigma^2_2(v^2_1+v^2_2)+\sigma^2_1\sigma^2_3(v^2_1+v^2_3)+\sigma^2_2\sigma^2_3(
v^2_2+v^2_3)}{\sigma_1\sigma_2\sigma_3}.\end{equation}
This expression must have a physical maximum since by perturbativity all parameters are below $4\pi$, and the existence of triangle inequalities for all $\rho_j$ implies a minimum for $\sigma_1\sigma_2 \sigma_3\neq0$.\\ 
The same sum is given in the $\jzs$ model by
\begin{equation}\mathrm{tr}\big(M^2_\tau\big)+\mathrm{tr}\big(M^2_\eta\big)
=2(\lambda_1v^2_1+\lambda_2v^2_2+\lambda_3v^2_3)+2\frac{\sigma^2_{31}\sigma
^2_{12}v^2_1+\sigma^2_{12}\sigma^2_{23}v^2_2+\sigma^2_{23}\sigma^2_{31}v^2_
3}{\sigma_{12}\sigma_{23}\sigma_{31}},\end{equation}with the same arguments
valid. As such, we conclude that there is no decoupling limit in the $\jzc$
and $\jzs$ models compatible with spontaneous CPV\footnote{The absence of a decoupling regime in the 2HDM with spontaneous CPV has been discussed in \cite{Nebot:2019lzf} (see also \cite{Faro:2020qyp}).}.

In this appendix, and throughout this paper, we have been using some param%
eters of the potential, alongside the masses and mixings of the scalars, w%
ith no concern for their relation, whose determination is beyond the scope 
of this paper. Nevertheless, we point out that there are 15 real parameters
in each potential, to be compared with seven masses (5 neutral and 2 charg%
ed scalars), twelve mixings (10 in the neutral and 2 in the charged sector%
), three vacuum expectation values $v_i$ and their two phase differences. Th%
us, we do not have the full freedom of a general 3HDM due to the symmetries
considered, but a rather constrained scenario with, for example, some vani%
shing mixing angles.

\clearpage

\providecommand{\href}[2]{#2}\begingroup\raggedright\endgroup

\end{document}